\documentclass[aps,pre,twocolumn,groupedaddress,showpacs]{revtex4}
\usepackage{graphicx}
\usepackage{dcolumn}
\usepackage{bm}
\usepackage{epsfig}
\usepackage{color}
\usepackage{float}
\input epsf
\epsfclipon
\newcommand{\argmin}{\arg\!\min}

\begin{document}
\title{Impact of lexical and sentiment factors on the popularity of scientific papers} 
\author{Julian Sienkiewicz and Eduardo G. Altmann}
\affiliation{Max Planck Institute for the Physics of Complex Systems, D-01187 Dresden, Germany}
\date{\today}
\begin{abstract}
We investigate how textual properties of scientific papers relate to the number of citations they receive. Our main finding is that correlations are non-linear and affect differently most-cited and typical papers. For instance, we find that in most journals short titles correlate positively with citations only for the most cited papers, for typical papers the correlation is in most cases negative. Our analysis of 6 different factors, calculated both at the title and abstract level of $4.3$ million papers in over 1500 journals, reveals the number of authors, and the length and complexity of the abstract, as having the strongest (positive) influence  on the number of citations. 
\end{abstract} 
\maketitle

\section{Introduction}
The number of citations an article receives can be considered a proxy for the attention or popularity the article achieved in the scientific community. Citations play a crucial role both in the evolution of science \cite{scfields,kuhn,scdyn,perc,sinatra} as well as in the bibliometric evaluation of scientists and institutions, in which case the number of citations is often tacitly taken as a measure of quality. Understanding which factors in a paper contribute or correlate with citations has been the subject of a number of investigations (see Refs.~\cite{didegah1,didegah2,onodera} for reviews). Diversity in the affiliation of authors, multinationality, multidisciplinarity, and number of references, figures or tables have all been identified as factors that positively correlate with citations. 

Here we perform a more systematic investigation of how different textual properties of scientific papers affect the number of citations they acquire (see Appendix \ref{methods:data} for data description). A classical result, which motivates our more general analysis, is the negative correlation between title length and citations (i.e., shorter the titles more citations)~\cite{paiva,wesel,rostami,letchford}. In our analysis we consider additionally the complexity and the sentiment of the text both at the title and the abstract, see table~\ref{tab:factors}. Lexical complexity is usually considered as proportional to the effort needed (by non-experts) to understand the texts. The three measures of text complexity we use (see table~\ref{tab:factors}) take into account the number of different words in the text (normalized by the length) and the length of these words in syllables (see Appendix \ref{methods:text} for details). In several previous studies authors used the concept of the sentiment analysis (i.e., emotional content) of the examined text/messages. In general, psychologists are able to specify several dimensions of emotions, reaching as far as 12~\cite{emo12}. However two of them --- {\it valence} and {\it arousal} --- are probably the best recognized and the most frequently used. Valence reflects the emotional sign of the message (negative, neutral, positive) while arousal is used to describe the level of activation (low, medium, high). Pairs of valence and arousal can indicate the specific emotion type~\cite{russel}, e.g., fear (negative and aroused), sad (negative and not aroused) etc., however they can be also utilized as independent variables. For example valence as a standalone dimension has successfully been used to detect collective states of online users~\cite{ania}, to indicate the end of online discussions~\cite{entropy} or to predict the dynamics of Twitter users during Olympic Games in London~\cite{janek}. Lately this kind of analysis has also been introduced to judge upon the role of negative citations~\cite{neg_cites}, citation bias~\cite{yu}, and to check what boosts the diffusion of scientific content~\cite{milkman}. Here we quantify arousal and valence through dictionary classifier, see Appendix \ref{methods:sent}. 

\footnotesize
\begin{table}[!h]
\begin{tabular}{llll}
  \hline
property & title & abstract \\ 
  \hline
{\bf length} & number of chars & number of words\\
\hline
{\bf complexity} &  --- & Gunning fog index $F$\\ 
 & $z$-index & $z$-index\\ 
 & Herdan's $C$ & Herdan's $C$\\ 
\hline
{\bf sentiment} & valence & valence \\
& arousal & arousal \\ 
\hline
{\bf number of authors} &  &  \\ 
\hline
\end{tabular}
\caption{List of textual factors whose relation to citations we investigate in our paper. Whenever possible, factors are obtained on the title and abstract of a paper.  See Appendices \ref{methods:text} and \ref{methods:sent} for exact definitions. Additionally, we consider the number of authors (motivated by previous studies e.g.,~\cite{didegah2,jose}).}
\label{tab:factors} 
\end{table}

\normalsize
\begin{figure*}
  \begin{tabular}{c}
	\includegraphics[width = \textwidth]{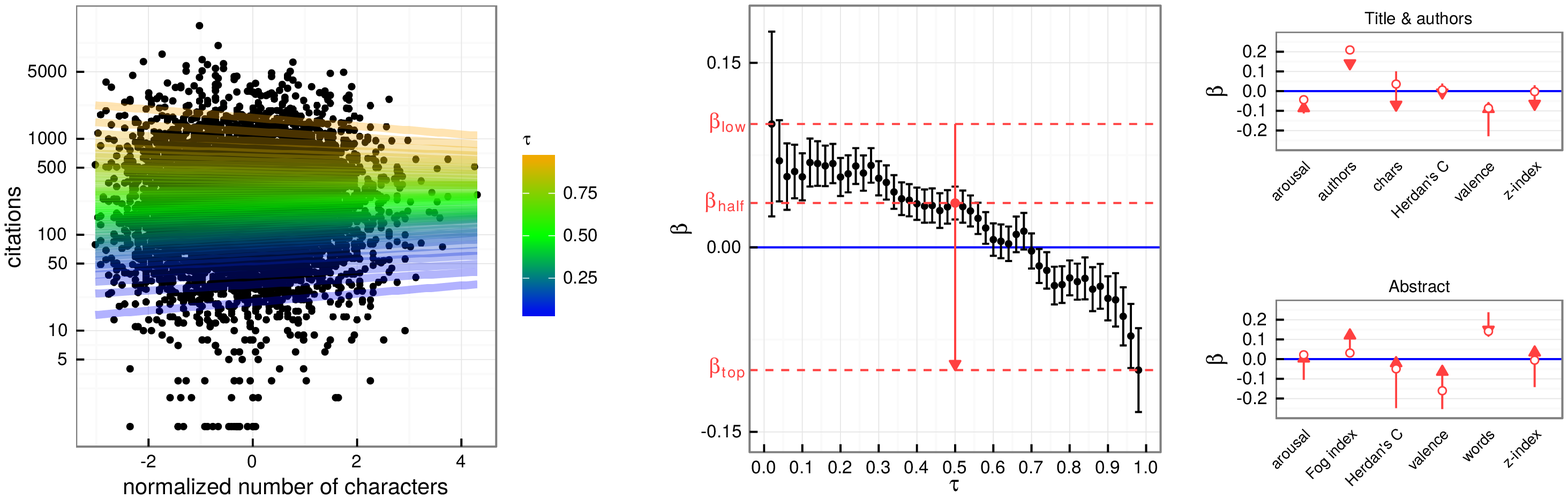}\\
	\includegraphics[width = \textwidth]{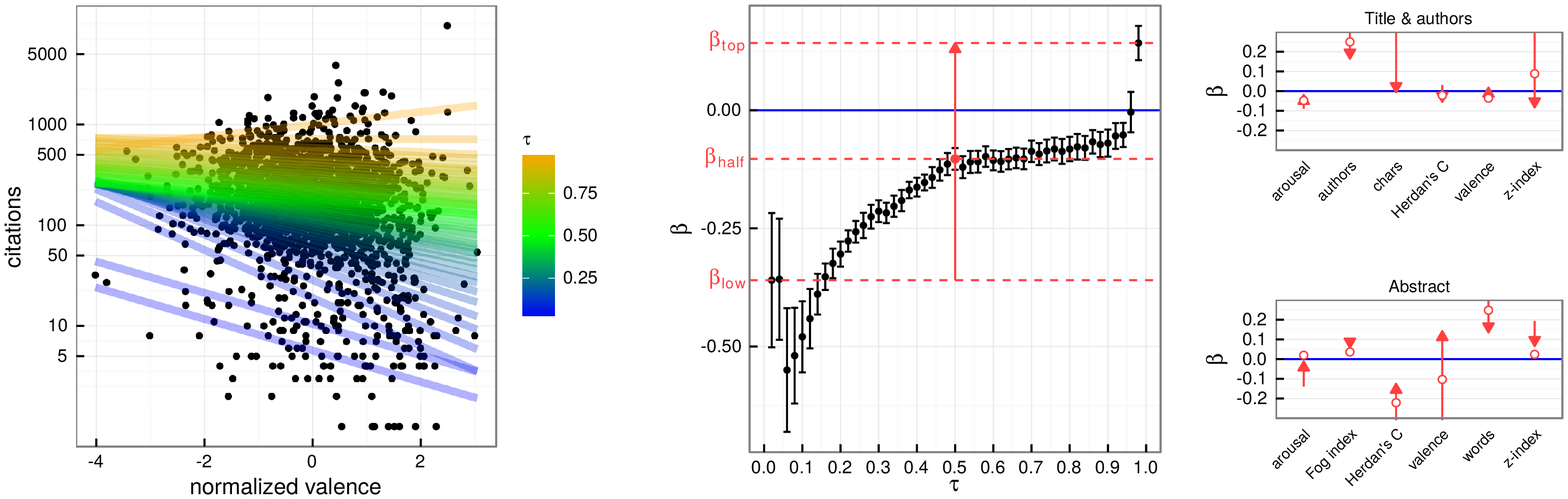}\\
  \end{tabular}
  \caption{Relation between different factors and the number of citations in two journals: \textit{Science} (top) and \textit{Nature Genetics} (bottom). Left-side plots: each black dot corresponds to one paper and lines show QR results for color-coded quantiles $\tau = \{0.02, 0.04,..., 0.98 \}$. Middle panels: $\beta$ coefficients (slopes of QR in the left panel) as a function of quantile $\tau$. The red arrows (\textit{summary pointers}) show $\beta_{\rm{low}} \equiv \beta(\tau = 0.02)$, $\beta_{\rm{half}} \equiv \beta(\tau = 0.5)$ and $\beta_{\rm{top}} \equiv \beta(\tau = 0.98)$, as, respectively, the nock, a circle on the shaft, and the head of the arrow. Right panels: summary pointers for all factors.}
  \label{fig:qr}
\end{figure*}

\section{Results}
We are interested in quantifying the relationship between $X$ -- a real number that quantifies for each paper one of the textual factors listed in table~\ref{tab:factors} -- and the logarithm of the number of citations $Y \equiv \ln(\text{citations}+1)$.  We standardize $X$ in order to be able to compare the different factors (see Appendix \ref{methods:stand}) and we use the citations provided by Web Of Science at the end of 2014 for papers published in 1995-2004\footnote{This guarantees that papers have at least $10$ years to gain citations. Equivalent, but much more noisy (due to drastically limited number of data points) results, are obtained when looking only at papers published in the same year.}. Exemplary results of the $X$ vs. $Y$ relationship for two factors in two journals are shown in the left part of figure \ref{fig:qr}. The broad scattering of the points shows that visual inspection fails even to detect whether the relation between $X$ and $Y$ is positive or negative.  The simplest (and widely used) approach is to perform an ordinary (least square) linear regression $Y = \alpha^\dagger + \beta^\dagger X$, in which case $\beta$ is related to the Pearson correlation coefficient $r$ as $\beta = r \sigma_Y / \sigma_X$ (in fact, due to standardization of variable $X$ in our case $\beta^\dagger$ is simply $\mathrm{cov}_{XY}$).  For the data in figure \ref{fig:qr}, this yields: $\beta^\dagger=0.020 \pm 0.011$ with $p > 0.05$ for title length in {\it Science} case and $\beta^\dagger=-0.21 \pm 0.03$ with $p < 0.001$ for valence in {\it Nature Genetics}. In other words, the second example shows a negative correlation between valence and citations while the first shows no clear correlation between the number of characters and citations (we cannot reject the null hypothesis of lack of linear dependence at 5\% significance level). We note that the analysis of Ref.~\cite{letchford}, which identified a negative correlation between title length and citation, was restricted only to the most cited papers. This difference in the conclusion regarding the role of title length and the large variability shown in the data motivates us to go beyond the above described computation of linear correlations, which relies on the (homoscedasticity) assumption of uniform  errors in the {\it whole} dataset.

\subsection{Quantile regression (QR)}
Quantile regression ~\cite{qrmain} gives the opportunity to track the relation between variables for different {\it parts} of the dataset. The simple question it addresses is: what are the coefficients $\alpha$ and $\beta$ of a linear relation $Y = \alpha(\tau) + \beta(\tau) X$ that divides the dataset so that a fraction $\tau$ of points lies below the line and the remaining part (1-$\tau$) above it (a precise formulation of QR is shown in Appendix \ref{methods:qr}). We thus obtain a sequence of values $\beta(\tau)$ which can be thought as the quantification of the relation between $X$ and $Y$ at the $\tau$ quantile. The QR is widely used in different fields~\cite{qr} and has lately been applied to predict future paper citation basing on their previous history, i.e., early citations as well as on the Impact Factor ~\cite{qrpred}.

The results in the center panels of figure~\ref{fig:qr} show a clear $\tau$ dependence of $\beta$, a signature of the non-linearity of correlations. For instance, the top panel shows that for low values of $\tau$ there is a positive correlation between number of characters in the tile and citations, while for high $\tau$ the correlation is reversed. This shows the limitations of the popularized message~\cite{titscience,titnature} following Ref.~\cite{letchford} that shorter titles lead to more citation. This only holds if you know in advance that your paper will be among the top cited papers (longer titles seem to be better, e.g., in order to avoid being among the least cited papers). Similar observations (with the opposite trend) are observed in the lower panel for valence --- the emotional polarity --- contained in the abstract of {\it Nature Genetics} articles. These examples show that even simple textual variables can have a mixed relation to the number of citations acquired by the papers of a given journal. 
We repeated the QR analysis for all factors in more than 1500 journals\footnote{We perform QR fitting for each journal independently because journals are known to play an important role in the number of citations a paper receives.}. In our discussion of our different findings below we focus on three characteristic values of $\beta$ which represent the low-cited ($\beta_{\rm{low}} \equiv \beta(\tau = 0.02)$), typical ($\beta_{\rm{half}} \equiv \beta(\tau = 0.5)$),  and top-cited ($\beta_{\rm{top}} \equiv \beta(\tau = 0.98)$) papers (graphically represented in the central and left panels of figure \ref{fig:qr} by a {\it summary pointer}, i.e., a red arrow with a circle). 

\begin{figure*}
	\includegraphics[width = .7\textwidth]{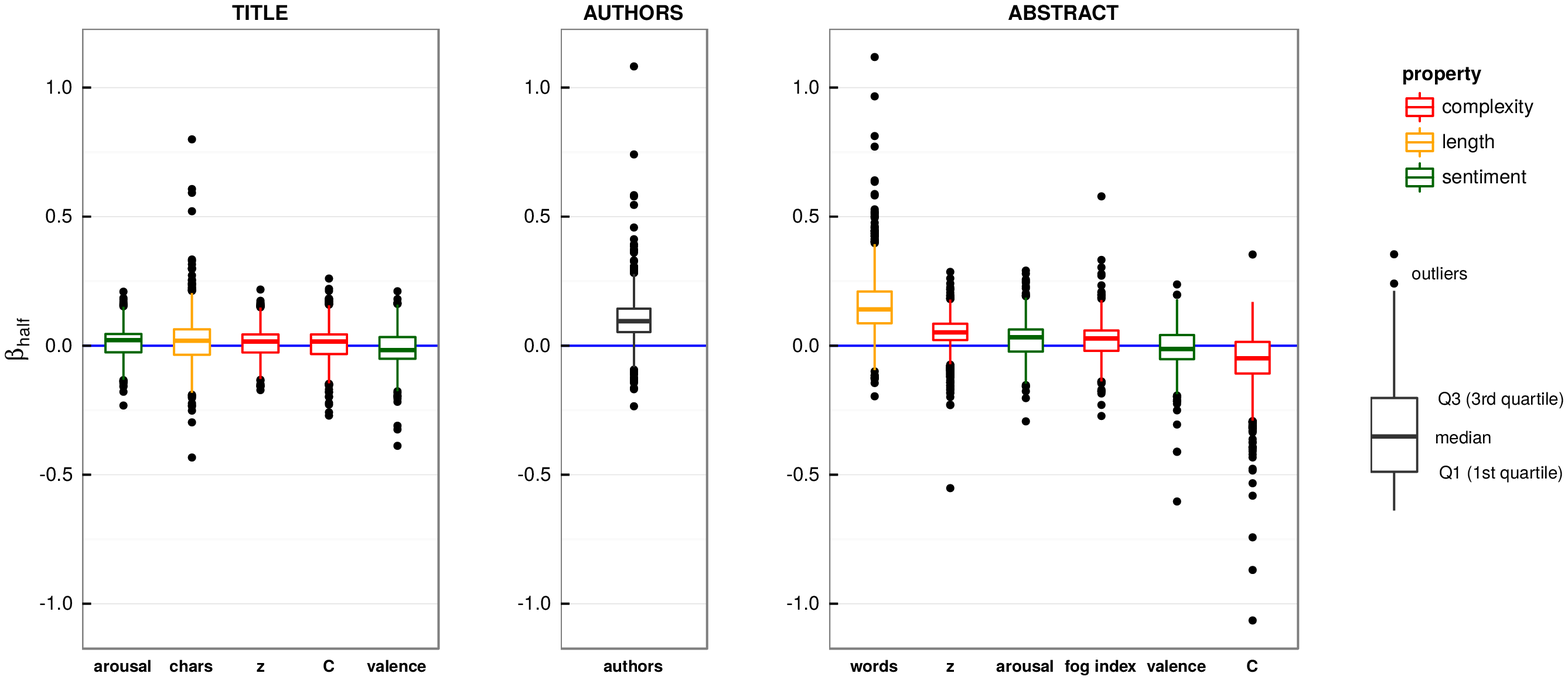}\\
	\caption{Strength of factors calculated over all journals. Box-plots (see definition on the right)  summarize the distribution of $\beta_{\rm{half}}$ values across different journals. Influential factors are identified as those for which $| \beta |$ is large for almost all journals (e.g., when the box does not contain $\beta_{\rm{half}} = 0$ line this implies that in at least 75\% of the journals the value of $\beta_{\rm{half}}$ is above or below zero).}
	\label{fig:box}
\end{figure*}

\subsection{Strength of factors}
In order to compare the strength of the effect of a factor on the number of citations we focus on the distribution of $\beta_{\rm{half}}$ (typical papers) across different journals. The linear relationship $Y = \ln(\rm{citations}) = \alpha + \beta X$ and the fact that $X$ is standardized imply that $\beta$ quantifies how much growth in citations should be expected from the variation of one standard deviation in one factor (e.g., $\beta=\ln 2 \approx 0.69$  means that the number of citations $Y$ doubles by moving one standard deviation in $X$).  Figure \ref{fig:box} summarizes the results and presents the factors ordered according to the median of the $\beta_{\rm{half}}$ distributions. The influence of factors is overall rather weak, as seen by the fact that for most journals $| \beta | <0.5$. Factors in the title are considerably weaker than those in the abstract or the one connected to the number of authors. The variation across journals is in general high, but higher in the title than in the abstract (possibly due to the fact that the estimations of $X$ are more robust in the abstract due to the larger amount of text). The strongest factors observed are: (i) the number of words in the abstract, (ii) the number of authors and (iii) $z$-index in the abstract. For those factors, over 75\% of journals (equivalently, the whole box) are placed above zero. The negative value of Herdan's $C$ can be attributed to its anticorrelation to the number of words (see Appendix \ref{methods:text}); when $C$ is accounted for that fact and presented in the form of $z$-index the value is positive. This means that for a typical paper and for most journals a more variable vocabulary (more unique words) translates into more citations. Similarly, the number of words in the abstract or the number of authors are positively correlated with the number of citations in almost all journals. 

\paragraph{Quantile dependence.} Now we quantify in which extent the influence of factors ($\beta$) varies across papers with different number of citations (the quantile $\tau$). We are particularly interested in the cases in which the effect of a given factor on the most successful papers is significantly different from the effect on typical papers. To quantify how typical this is, we count the number of journals for which $\beta_{\rm{top}} \neq \beta_{\rm{half}}$ is observed beyond the estimated uncertainties $\sigma_{\beta_{\rm{top}}}$, $\sigma_{\beta_{\rm{half}}}$, i.e., $| \beta_{\rm{top}} - \beta_{\rm{half}} | > (\sigma_{\beta_{\rm{top}}} + \sigma_{\beta_{\rm{half}}})$.  The results shown in table~\ref{tab:diff} reveal that overall this happens in about $1/3$ of the cases (it is more typical for text length and less typical for sentiment factors). Table~\ref{tab:diff} reveals also the factors for which $\beta_{\rm{top}} \neq \beta_{\rm{half}}$ because $\beta(\tau)$ grows in most journals (and thus $\beta_{\rm{top}} > \beta_{\rm{half}}$, as in the case of valence in the abstract), decays in most journals (and thus $\beta_{\rm{top}} < \beta_{\rm{half}}$, as in the case title length), or shows a mixed behavior in different journals (as in the case of arousal).

The next question we investigate is the extent into which the quantile dependence leads to a reversal of the effect of factors, i.e., when $\beta(\tau)$ crosses $0$. Table~\ref{tab:pos} shows the percentage of journals with positive $\beta_{\rm{low}}$, $\beta_{\rm{half}}$, and $\beta_{\rm{top}}$ coefficients for each factor. It shows that except for singular cases (marked by asterisk) the observations tend to be significantly different from chance ($50\%$). The variation across the different $\beta$'s (quantiles) quantifies the number of journals for which $\beta(\tau)$ crosses zero. 
Such a behavior has already been discussed for title length in $Science$ (see figure~\ref{fig:qr}), and table~\ref{tab:pos} confirms the generality of this observation (it shows for title length 72\% of journals with positive  $\beta_{\rm{low}}$ as compared to nearly 75\% with negative $\beta_{\rm{top}}$). In case of three factors (title length, Herdan's $C$ in the abstract, and valence in the abstract), we observe that moving form $\beta_{\rm{low}}$ to $\beta_{\rm{top}}$ we cross $50\%$, which indicates that for a certain range of $\beta$ the factor in question increases citations for most journals while for other $\beta$'s the opposite effect is typical across journals. 

\footnotesize
\begin{table*}
\begin{tabular}{lrccc}
\hline
property & factor & $\beta_{\rm{top}} > \beta_{\rm{half}}$ & $\beta_{\rm{top}} < \beta_{\rm{half}}$ & $\beta_{\rm{top}} \neq \beta_{\rm{half}}$\\
\hline
length & no. of characters (title) & 2.6\% & 44.4\% & 47.0\%\\
 & no. of words (abstract) & 8.3\% & 29.4\% & 36.7\%\\
 &  &  & mean & 41.9\%\\
\hline
complexity & Herdan's $C$ (title) & 18.7\% & 8.5\% & 27.2\%\\
& Herdan's $C$ (abstract) & 34.9\% & 6.5\% & 41.4\%\\
& $z$-index (title) & 8.3\% & 16.7\% & 25.0\%\\
& $z$-index (abstract) & 24.6\% & 7.7\% & 32.3\%\\
& fog index (abstract) & 26.4\% & 8.0\% & 34.4\%\\
 &  &  & mean & 32.0\%\\
\hline
sentiment & arousal (title) & 11.0\% & 13.5\% & 24.5\%\\
& arousal (abstract) & 15.7\% & 13.7\% & 29.4\%\\
& valence (title) & 16.1\% & 11.3\% & 27.4\%\\
& valence (abstract) & 29.2\% & 5.7\% & 34.9\%\\
 &  &  & mean & 29.1\%\\
\hline
& no. of authors & 4.0\% & 39.6\% & 43.6\%\\
\hline
 &  &  & overall mean & 33.7\%\\
\end{tabular}
\caption{Factors often affect top and typical papers differently. Percentage of journals for which $\beta_{\rm{top}} \neq \beta_{\rm{half}}$ are reported. The right column, $\beta_{\rm{top}} \neq \beta_{\rm{half}}$, is the sum of the two others.}
\label{tab:diff}
\end{table*}

\begin{table}
\begin{tabular}{lrccc}
\hline
property & factor & $\beta_{\rm{low}} > 0$ & $\beta_{\rm{half}} > 0$ & $\beta_{\rm{top}} > 0$\\
\hline
length & no. of characters (title) & 71.4\%  & 56.2\% & 27.7\%\\
 & no. of words (abstract) & 96.5\% & 96.7\% & 83.4\%\\
\hline
complexity & Herdan's $C$ (title) & 50.1\%$^{*}$ & 56.7\% & 62.4\% \\
& Herdan's $C$ (abstract) & 19.4\% & 28.1\% & 51.2\%$^{*}$ \\
& $z$-index (title) & 62.0\% & 58.2\% & 47.9\%$^{*}$ \\
& $z$-index (abstract) & 71.3\% & 82.1\% & 81.0\% \\
& fog index (abstract) & 62.9\% & 68.2\% & 72.9\% \\
\hline
sentiment & arousal (title) & 56.5\% & 61.6\% & 58.3\% \\
& arousal (abstract) & 62.8\% & 67.8\% & 61.9\% \\
& valence (title) & 42.9\% & 43.1\% & 49.5\%$^{*}$ \\
& valence (abstract) & 42.0\% & 47.6\%$^{*}$ & 63.4\% \\
\hline
& no. of authors & 93.4\% & 92.5\% & 61.6\% \\
\hline
\end{tabular}
\caption{Percentage of journals with positive $\beta_{\rm{low}}$, $\beta_{\rm{half}}$, and $\beta_{\rm{top}}$  for each factor. All values are statistically significant ($p < 0.001$) except for those marked with an asterisk $^{*}$ (see Appendix \ref{methods:stat}).}
\label{tab:pos}
\end{table}

\normalsize
The combination of the results of these two tables allows for a more complete picture of the $\tau$ dependence on $\beta$ for different factors. For instance, the number of authors and the number of characters in the title can be identified as the ones that exhibit the strongest systematic trend of decaying $\beta(\tau)$ (in about 40\% of journals, as shown in table~\ref{tab:diff}). However, only for the number of authors the majority of the values are above zero (see table~\ref{tab:pos}), i.e., the value of $\beta$ for top papers is less then for typical ones but it still stays positive. On the other hand, in the case of the number of characters not only is $\beta$ smaller for top papers as compared to typical ones but it changes its sign as well.  Sentiment factors (except for the valence in the abstract) bring no overall information about the trend --- the number of up- and downward occurrences is similar. Notably, there is a strong coincidence between $z$-index and fog index in the abstract, suggesting that although those two quantities have different definitions, both indicate the increase of correlations between abstract complexity and citations.

\begin{figure*}
\begin{tabular}{cc}
\includegraphics[width=.9\textheight,angle=90]{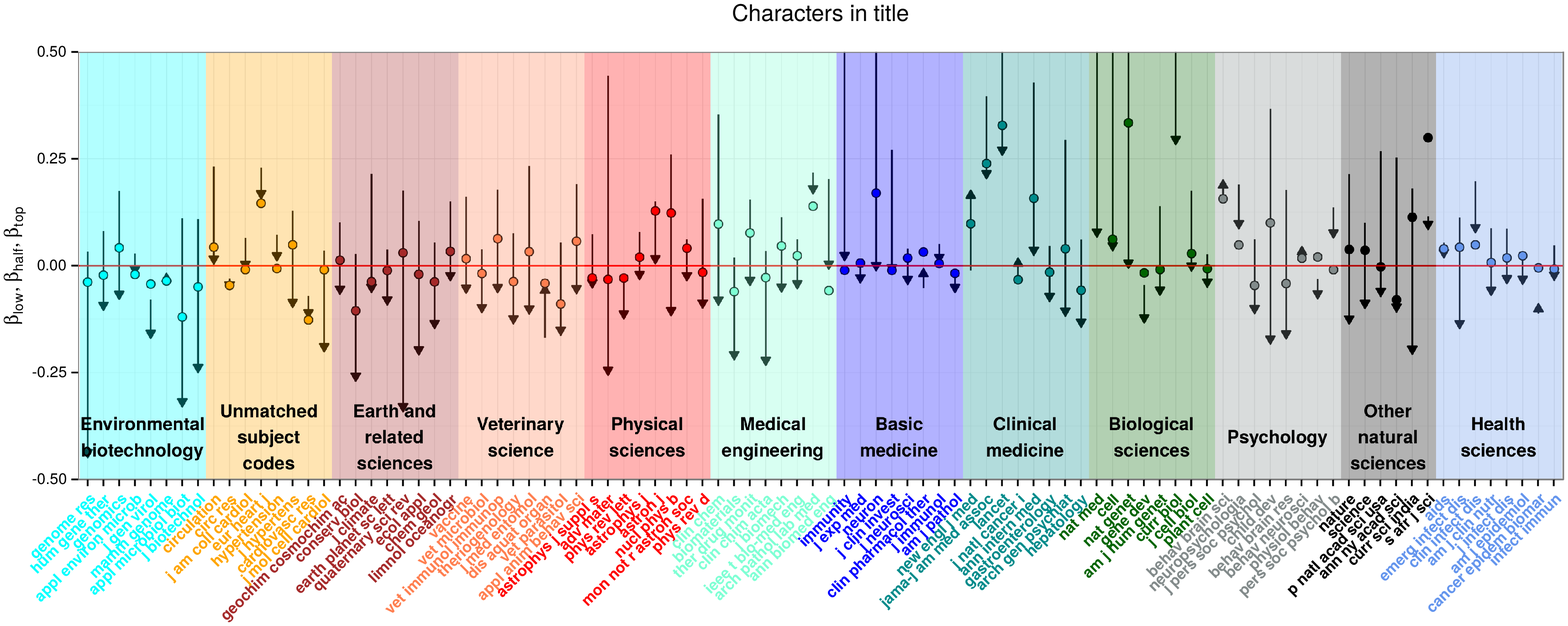} & \includegraphics[width=.9\textheight,angle=90]{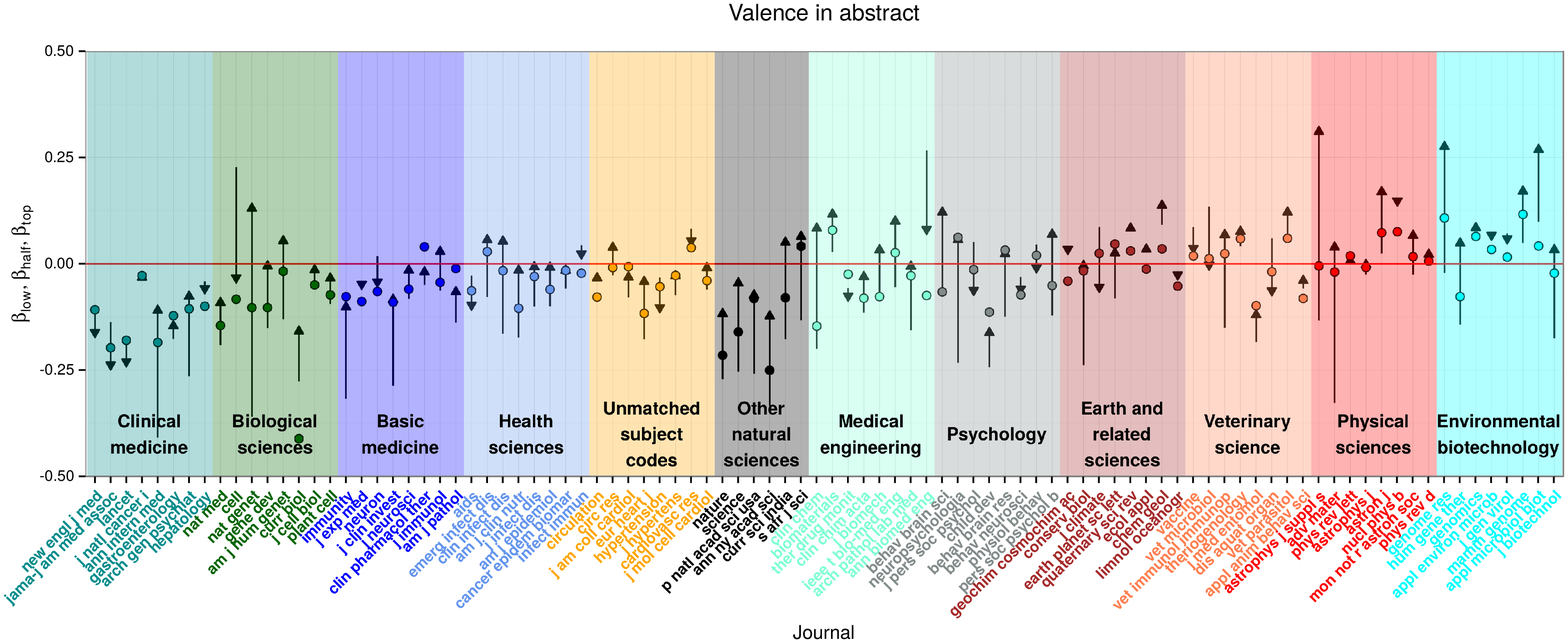}\\
\end{tabular}
\caption{Summary pointers showing $\beta_{\rm{low}}$, $\beta_{\rm{half}}$, and $\beta_{\rm{top}}$ for two factors number of title characters (top) and valence in abstract (bottom) (see figure~\ref{fig:qr} for the definition of summary pointers). Journals are grouped according to the OECD bibliographic categories \cite{oecd}. The 8 journals with highest Impact Factor in each category are shown ($6$ for \textit{Other natural sciences}). The categories are sorted with respect to the number of positive $\beta_{\rm{half}}$ values. Testing null hypothesis that categories are randomly attributed to journals (we compare the average standard deviation within categories with a random attribution of categories to journals) yields p-values $p=0.002$ for title length and $p < 10^{-8}$ for valence in abstract. The same procedure performed for Impact Factor (by creating 12 categories according to decreasing IF) gives $p=0.02$ for title length and $p < 10^{-5}$ for valence in abstract, suggesting higher impact exerted by scientific category.}
\label{fig:journ}
\end{figure*}

{\bf Variability across journals.} 
The  large variability across journals apparent in all our analysis can have different origins. One possibility is that certain journals are read only by specific (scientific) communities. To address that issue, in figure \ref{fig:journ} we group the journals in disciplines according to their OECD subcategory~\cite{oecd} and show summary pointers (introduced in figure~\ref{fig:qr}) for two factors. The results indicate that the variation across journals is partially explained by disciplines, e.g., for {\it Clinical medicine} all values of $\beta$ in case of valence in abstract are below zero while for {\it Physical sciences} the majority is positive. Another possibility is that more popular journals are different than less popular journal. To address this option, journals inside each discipline in figure~\ref{fig:journ} are ranked by their Impact Factor index (IF). No clear tendency can be visually identified, however by comparing with a random attribution of IF, popularity proves to be statistically significant, although to much less extent than scientific discipline (see caption for figure~\ref{fig:journ}). Figure ~\ref{fig:journ} allows also for a straightforward comparison of the strength of title length and abstract valence factors in different journals. By calculating $\exp(\beta \Delta X)$ one can directly estimate how much gain in citations is obtained on average by a move in $\Delta X$ standard deviations in the variable $X$ (e.g., for title length in the journal \textit{Lancet} $\beta_{\rm{half}}=0.33$ and thus extending the length of the title by one standard deviation gives almost 40\% gain in citations; for \textit{Nature}, $\beta_{\rm{half}}=0.038$ and thus one obtains less than 4\% gain).

\section{Discussion and Conclusion}
In this paper we investigate the {\it importance} of  factors of scientific papers on the popularity they acquire. As factors we consider the number of authors of the paper and also text-related properties that quantify the length of title and abstract, the complexity of the vocabulary, and sentiment based on the used words. These factors capture different stylistic dimensions of scientific writing and were selected also based on previous works that indicated a correlation to the number of citations. We found that the factors with a stronger (positive) effect on citations are the number of authors and the length of the abstract. Text complexity is positive correlated with citation at the level of the abstract, while we could not detect a strong effect within the title. The agreement of two factors designed to quantify text complexity -- the $z$-index and Gunning fog index -- support this conclusion (the opposite result is obtained if Herdan's C measure is used, but we attribute this to the negative correlation of this measure with text length). In terms of the sentiment factors, the level of arousal a title or abstract invokes is poorly correlated with citations. This result should be examined more carefully as there are controversies as to the relation between text polarity and information contained therein (see \cite{garcia,dodds} and the following discussion). Also the vocabulary on which we rely in this study \cite{warriner} has been obtained by evaluating the common reception of words. This fact can strongly affect the value of valence, e.g., a highly negative word ``cancer'' in medical papers.

The discussion above, and the fact that a statistically significant effect is present for most factors, should not hide that the effect is typically weak ($|\beta| < 0.5$ for most factors, quantiles $\tau$, and journals) and that there are strong fluctuations across papers and journals. For instance, a positive correlation between number of characters and citations for {\it all} the quantiles is measured in the {\it New England Journal of Medicine}, while a negative correlation is observed in the overwhelming majority of other journals. One of the main findings of our paper is that the factors vary also strongly depending whether the analysis uses all or only the most cited papers. We quantified this effect by the  dependence of $\beta$ on the quantile $\tau$ in a Quantile Regression analysis. One example in which this effect is particularly strong is the role of title length in figure \ref{fig:qr}. In the public media~\cite{titscience,titnature}, the message behind the finding~\cite{letchford} of negative correlation between text length and citations was that authors should write shorter titles to achieve more citations. While this simple message is appealing and agrees with some stylistic recommendations, our results show that for most journals this is wrong (even if one assumes that there is a causal relation behind the correlations). The negative correlation is found only in the most cited journals, for typical journals the correlation is positive (longer titles are better). This suggests that papers with short titles show a larger variation on the number of citations and can be very well cited or very poorly cited. A similar behavior is observed in other factors, and a significant dependence on $\tau$ is seen on average in $1/3$ of the journals. 

Altogether, our results indicate that textual properties of title and abstract have non-trivial effects in the processes leading to the attribution of citations. In particular, the effect varies significantly between papers with usual number of citations and with large number of citations. This finding is even more important considering that the number of citations across papers varies dramatically. The weak signal we detect can be considered also a sign that the quantities we measure have limited information, e.g., expressing the impact of publications by single number (the number of citations) can be misleading and lacking information (a point that has been previously raised, e.g., in Ref. \cite{bollen}). The overall estimates (calculated over a set of journals or categories) may dim the clear picture one receives while observing a specific journal. For authors interested in how to write the title and abstract of their paper, we recommend looking at the values of $\beta_{\rm{half}}$ and $\beta_{\rm{top}}$ of the different factors for the specific journals of interest.

\begin{acknowledgments}
We thank Margit Palzenberger and the Max-Planck-Digital-Library for providing access to the dataset used in this paper.
\end{acknowledgments}

\appendix
\section{Data}\label{methods:data}
We obtained data from the {\it Web of Science} service about the papers marked as ``articles'' published in the period of 1995 --- 2004 that fulfil the following two conditions: (i) the journal where the article has been published had to be active in all the mentioned years and (ii) there had to be at least $1000$ articles published in total in this journal in the given period. By applying this filtering we obtained over $4~300~000$ articles from over $1500$ different journals containing information about the title of the paper, the number of its authors, full abstract contents and OECD category it had been classified to. Additionally, for each of the record we also recorded the number of citations it acquired between being published and 31st December 2014. Data processing, plots and statistical analysis have been performed using \textsf{R} language \cite{R}.\\

\section{Text properties}\label{methods:text}
The most obvious candidates for quantitative factors that could be used to describe the paper are the number of words or the number of characters. In the case of the title the second option has been used while in the case of abstract --- the first one. Additionally the number of authors has also been used as in a previous study it had been shown to be an important factor \cite{jose}. As it concerns the complexity of the vocabulary a way to account for that is to measure a so-called Herdan's C index (see e.g., \cite{herdan} p. 72), defined for each paper $i$ as
\begin{equation}
C_i = \frac{\log N_i}{\log M_i},
\end{equation}
where $M_i$ stands for the text length (number of words) and $N_i$ is the vocabulary size (i.e., the number of unique words) of paper $i$. To overcome methodological shortcomings of this traditional approach (e.g., no fluctuations effect included)  it has recently been proposed \cite{gerlach} to use a $z$-score that shows how much the obtained pair ($N_i$, $M_i$) is different form the expected value $\mu(M)$ in units of standard deviations $\sigma(M)$
\begin{equation}
z_{N_i, M_i} = \frac{N_i - \mu(M_i)}{\sigma(M_i)},
\end{equation}
where $\mu(M_i)$ and $\sigma(M_i)$ were obtained empirically using all papers in our database.
Finally one might also take into account the complexity of the used words. A classical quantity to measure this effect is so-called ``Gunning fog index''  $F_i$ \cite{gunning}, defined for each paper $i$ as
\begin{equation}
F_i = 0.4 \left( \frac{\# words_i}{\# sentences_i} + 100\frac{\# complexwords_i}{\# words_i} \right),
\end{equation}
where a complex a word is a one that has more than two syllables \cite{syl}. Fog index is widely used as its value can be connected to the number of formal years of education needed to understand the text at first reading. Because of the absence of sentences fog index has not been calculated in case of title (i.e., a typical title contains only one sentence therefore $F_i$ is highly correlated with the number of words).\\

\section{Sentiment properties}\label{methods:sent}
In study the idea of a {\it dictionary emotional classifier} has been used: in this approach one takes the dictionary of words that had been tagged for valence and arousal and calculates the mean arithmetic value of all the recognized words. Thus in case of each paper we have separately valence (and arousal) values for title and abstract. We have used a very recent study \cite{warriner} which contains norms for almost 14.000 English words, where valence ($v$) and arousal ($a$) are given as real numbers in the scale $[1; 9]$ (i.e., $v$ below 5 is negative, while $v > 5$ means positive words, low $a$ values indicate low arousal, while high $a$ is high arousal). The total valence and arousal were obtained as the average of all words in the title or abstract.

\section{Standardization}\label{methods:stand}
In order to make comparison among different factors each factor $x$ has been separately standardized with respect to journal, i.e. for each $i$
\begin{equation}
\hat{x}_i = \frac{x_i - \mu(x)}{\sigma(x)},
\end{equation}
where $\mu(x)$ and $\sigma(x)$ are, respectively, sample mean and variance over factor $x$ in a journal $j$ it belongs to.

\section{Quantile regression}\label{methods:qr}
In the approach of quantile regression \cite{qrmain,qr}, having $k$ factors (variables) $X_k$ and an observable $Y$, we are able to obtain a regression line defined by coefficients $\beta_i(\tau)$
\begin{equation}
Y = \beta_0(\tau) + X_1 \beta_1(\tau) + ... + X_k \beta_k(\tau) =\mathbf{X} \mathbf{\beta}
\end{equation}
for a given quantile $\tau$ by solving the minimization problem
\begin{equation}
\hat{\beta}(\tau) = \argmin_{\beta \in R^k} \sum_{i=1}^n \left( \rho_{\tau}(Y_i - \mathbf{X} \mathbf{\beta}) \right),
\end{equation}
where $\rho_{\tau}(y) = |y(\tau - \mathcal{I}_{(y < 0)})|$ is called loss function ($\mathcal{I}$ is indicator variable). In this study we restrict ourselves to case where
\begin{equation}
\ln Y = \alpha(\tau) + X \beta(\tau),
\end{equation}
i.e., we examine the influence of each of the factors separately. As the logarithm is an increasing function, the logarithm of the $p$-th quantile is equal to the $p$-th quantile of the log-transformed citation counts. For computational purposes we used \textsf{R}'s \textit{quantreg} package \cite{R_QR}.

\section{Statistical analysis}\label{methods:stat}
We test if the number of positive values of $\beta_{\rm{low}}$, $\beta_{\rm{half}}$, and $\beta_{\rm{top}}$ is significantly different from the one obtained by chance (i.e., by randomly choosing ``+'' or ``-'' signs with equal probability $q=1/2$). This statistics follows a binomial distribution. However, as the number of samples (journals) $n$ is large ($n > 1500$) we simply use normal distribution $N(\mu, \sigma)$, with $\mu = nq$ and $\sigma = \sqrt{nq(q-1)}$ with $q = 1/2$. We consider the observation to be statistical significant at if the measured number of positive $\beta$ differs from $\mu$ by more than $3\sigma$  (i.e., p-value is less than $0.001$).

\end{document}